\documentclass[aps,prl,twocolumn]{revtex4-1}
\usepackage{amsmath}
\usepackage{graphicx}
\usepackage{esint}
\usepackage{epstopdf}%This line makes .eps figures into .pdf - please comment out if not required.
\graphicspath{{pict/}{./}}
\usepackage{bm}%
\newcounter{Fig}

\newcommand{\be}{\begin{equation}}
\newcommand{\ee}{\end{equation}}

\begin{document}

%\title{Q-factor enhancement for all-dielectric nanoresonators through relaxed total internal reflection}
\title{Generalized magnetic mirrors}
\author{Wei Liu}
\email{wei.liu.pku@gmail.com}
%\footnote{wei.liu.pku@gmail.com}
\affiliation{College of Optoelectronic Science and Engineering, National University of Defense
Technology, Changsha, Hunan 410073, P. R. China}
%\author{TBD}
%\affiliation{Nonlinear Physics Centre,  Research
%School of Physics and Engineering, Australian National University,
%Canberra, ACT 0200, Australia}
%\author{Andrey E. Miroshnichenko}
%\affiliation{Nonlinear Physics Centre,  Research
%School of Physics and Engineering, Australian National University,
%Canberra, ACT 0200, Australia}
%\author{Yuri S. Kivshar}
%\affiliation{Nonlinear Physics Centre,  Research
%School of Physics and Engineering, Australian National University,
%Canberra, ACT 0200, Australia}
%\affiliation{Department of Nanophotonics and Metamaterials, ITMO University, St. Petersburg 197101, Russia}
%\pacs{
%        78.67.-n,   % Optical properties of low-dimensional, mesoscopic, and nanoscale materials and structures
%        42.25.Fx,   % Diffraction and scattering
%%        73.20.Mf,   % surface and interface excitations,
%        78.67.Pt   % Multilayers; superlattices; photonic structures; metamaterials (see also 81.05.Xj, Metamaterials for chiral, bianisotropic and other complex media)
%%%       42.25.Bs    % Wave propagation, transmission and absorption [see also 41.20.Jb�in electromagnetism; for propagation in atmosphere, see 42.68.Ay; see also 52.40.Db Electromagnetic (nonlaser) radiation interactions with plasma and 52.38-r Laser-plasma interactions�in plasma physics]
%%%       42.25.-p    % Wave optics,
%}

\begin{abstract}
%We propose generalized magnetic mirrors that can be achieved by excitations of sole electric resonances. Conventional approaches to obtain magnetic mirrors rely heavily on exciting the fundamental magnetic dipoles, whereas here we reveal that besides magnetic resonances, electric resonances of higher orders can be also employed to obtain highly efficient magnetic mirrors. Based on the electromagnetic duality, it is also shown that electric mirrors can be achieved by exciting  magnetic resonances. We provide direct demonstrations of the generalized mirrors proposed in a simple system of one-dimensional periodic array of all-dielectric wires. This may inspire many investigations aiming to realise, replying on sole electric resonances, various advanced functionalities that were originally believed unobtainable without magnetic responses.
We propose generalized magnetic mirrors that can be achieved by excitations of sole electric resonances. Conventional approaches to obtain magnetic mirrors rely heavily on exciting the fundamental magnetic dipoles, whereas here we reveal that besides magnetic resonances, electric resonances of higher orders can be also employed to obtain highly efficient magnetic mirrors. Based on the electromagnetic duality, it is also shown that electric mirrors can be achieved by exciting  magnetic resonances. We provide direct demonstrations of the generalized mirrors proposed in a simple system of one-dimensional periodic array of all-dielectric wires, which may shed new light to many
advanced fields of photonics related to resonant multipolar excitations and interferences.
\end{abstract}
\maketitle

Displacement currents induced artificial magnetic responses, especially magnetic dipoles (MDs), serve as the cornerstone of many branches of metamaterials~\cite{Pendry1999_ITMT,Linden2004_science,Cai2010_book}, and play a central role in lots of exotic optical and photonic phenomena~\cite{Cai2010_book,Soukoulis2011_NP,Zheludev2012_NM}. Besides the relatively complex meta-atoms of designed shapes, it is recently discovered that optically-induced magnetism exists within simple high-index dielectric particles, which stems from magnetic Mie resonances supported~\cite{Zhao2009_materialtoday,jahani_alldielectric_2016,KUZNETSOV_Science_optically_2016}. Consequently, a new branch of all-dielectric meta-photonics emerges, which relies on a full exploitation of excitations and interferences of both electric and magnetic resonances~\cite{Liu2014_CPB,SMIRNOVA_Optica_multipolar_2016,LIU_ArXivPrepr.ArXiv160901099_multipolar_2016}, and enables much more flexible control of both magnitudes and phases of the fields~\cite{jahani_alldielectric_2016,KUZNETSOV_Science_optically_2016,Liu2014_CPB,SMIRNOVA_Optica_multipolar_2016,LIU_ArXivPrepr.ArXiv160901099_multipolar_2016,STAUDE_NatPhoton_metamaterialinspired_2017}. Moreover, such a field rapidly hybridizes with other branches of metasurfaces, optical nanoantennas and topological photonics~\cite{SLOBOZHANYUK_NatPhoton_threedimensional_2016}, which renders enormous extra freedom for manipulations of various kinds of light-matter interactions and makes possible lots of novel photonic functionalities and devices~\cite{CHEN_Rep.Prog.Phys._review_2016,GENEVET_OpticaOPTICA_recent_2017-1,DING_ArXiv170403032Phys._gradient_2017}.

A recent rather remarkable achievement based on optically-induced magnetism is magnetic mirrors (MMs) obtained at various spectral regimes~\cite{SCHWANECKE_J.Opt.PureAppl.Opt._optical_2006,ESFANDYARPOUR_Nat.Nanotechnol._metamaterial_2014,LIU_OpticaOPTICA_optical_2014,MOITRA_ACSPhotonics_largescale_2015,CHOI_Adv.Opt.Mater._nearzero_2015,
HEADLAND_Adv.Mater._terahertz_2015,LIN_Appl.Phys.Lett._dielectric_2016,MA_ACSPhotonics_terahertz_2016-1,SONG_Conf.LasersElectro-Opt.2016Pap.JW2A27_silicon_2016,ZHOU_Opt.Lett.OL_proposal_2016}. MMs can effectively eliminate the half-wave loss of electric fields that exists at the conventional medium interfaces, and thus have the electric fields enhanced close to the boundaries. This is of great importance for various applications including sensing, imaging, and many other applications related to light-matter interaction enhancement~\cite{SCHWANECKE_J.Opt.PureAppl.Opt._optical_2006,ESFANDYARPOUR_Nat.Nanotechnol._metamaterial_2014,LIU_OpticaOPTICA_optical_2014,MOITRA_ACSPhotonics_largescale_2015,CHOI_Adv.Opt.Mater._nearzero_2015,
HEADLAND_Adv.Mater._terahertz_2015,LIN_Appl.Phys.Lett._dielectric_2016,MA_ACSPhotonics_terahertz_2016-1,SONG_Conf.LasersElectro-Opt.2016Pap.JW2A27_silicon_2016,ZHOU_Opt.Lett.OL_proposal_2016}. Nevertheless, similar to many novel functionalities obtained in metamaterials and metasurfaces~\cite{Cai2010_book,Soukoulis2011_NP,Zheludev2012_NM,Zhao2009_materialtoday,SLOBOZHANYUK_NatPhoton_threedimensional_2016,jahani_alldielectric_2016,KUZNETSOV_Science_optically_2016,Liu2014_CPB,SMIRNOVA_Optica_multipolar_2016,LIU_ArXivPrepr.ArXiv160901099_multipolar_2016,STAUDE_NatPhoton_metamaterialinspired_2017,CHEN_Rep.Prog.Phys._review_2016,GENEVET_OpticaOPTICA_recent_2017-1,
DING_ArXiv170403032Phys._gradient_2017}, the existing approaches to obtain MMs heavily rely on the excitations of the fundamental MDs~\cite{LIU_OpticaOPTICA_optical_2014,MOITRA_ACSPhotonics_largescale_2015,LIN_Appl.Phys.Lett._dielectric_2016,MA_ACSPhotonics_terahertz_2016-1,SONG_Conf.LasersElectro-Opt.2016Pap.JW2A27_silicon_2016}. Then it is vital to ask: for the demonstration of both MMs and many other exotic phenomena, are the magnetic responses (especially MDs) really essential? Can the magnetic resonances be replaced by sole electric ones  (or vice versa) to realize identical functionalities?

\begin{figure}[tp]
\centerline{\includegraphics[width=8.9cm]{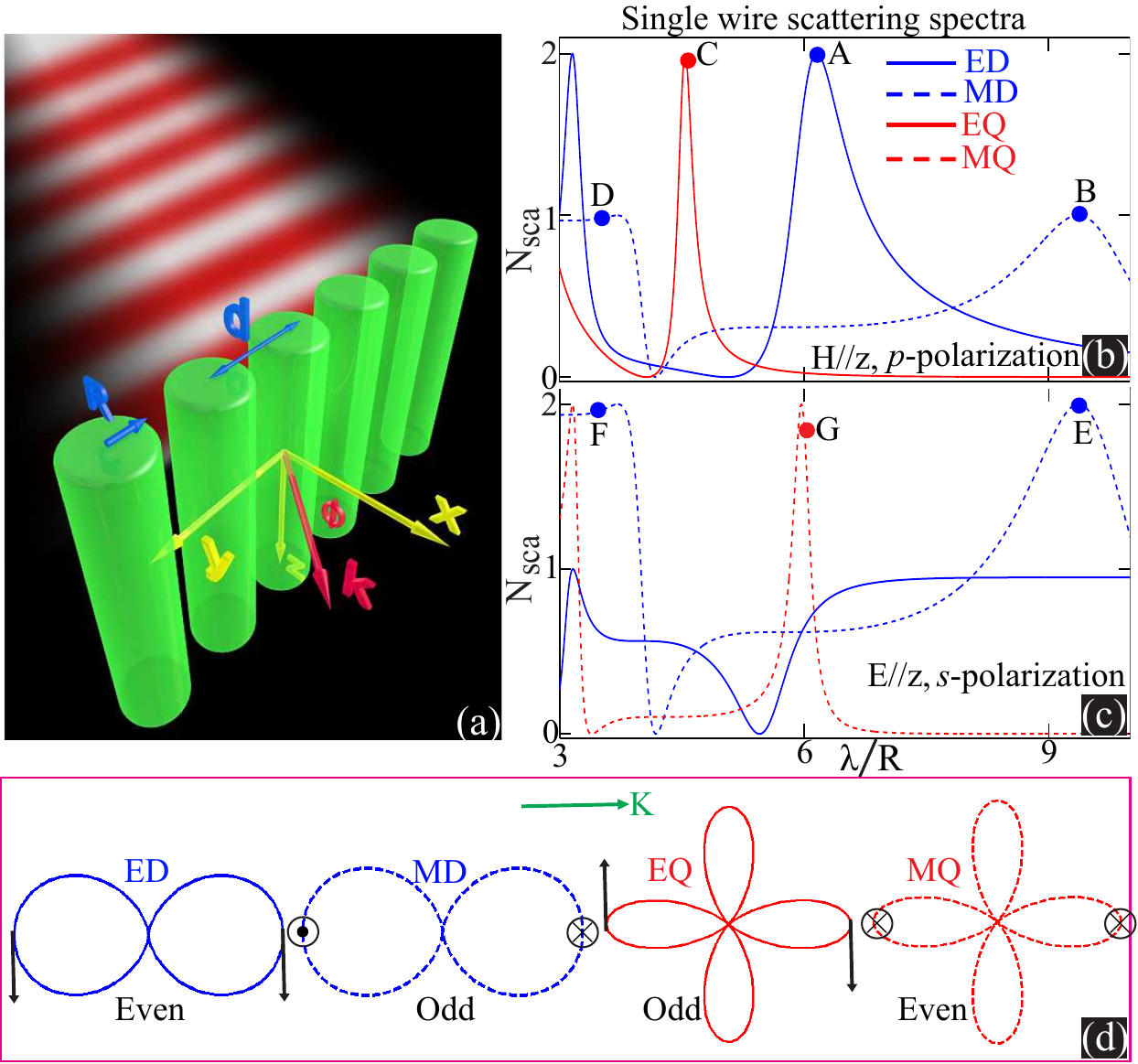}} \caption {\small (a) Schematic of a normally incident plane wave (with wave-vector $\mathbf{k}$ on the $\textbf{x}-\textbf{y}$ plane and the polar angle is $\phi$) shining on a 1D array of identical wires of refractive index $n$, radius $R$ and unit-cell size $d$. The wave can be $\textbf{\textit{p}}$- or $\textbf{\textit{s}}$-polarized. (b) and (c) show the multipolar scattering spectra for an individual wire of $n=3.5$ with $\textbf{\textit{p}}$- and $\textbf{\textit{s}}$-polarized waves respectively. The solid and dashed curves correspond to electric and magnetic multipoles respectively. Points A-G (related data specified in main text) at or very close to the central resonant positions of different multipoles are indicated. (d) Schematic illustrations of the scattering parities for the first four multipoles. The exciting field is from the left, and the black arrows (both in plane and perpendicular to the plane ones) indicate the scattered electric fields are in-phase ($\uparrow$ and $\odot$) or out-of-phase ($\downarrow$ and $\otimes$) with respect to the incident electric field $\mathbf{E_0}$ at the forward or backward directions.}\label{fig1}
\end{figure} %($\lambda_{A-G}/R=6.16,~9.54,~4.58,~3.54,~9.54,~3.54$, and $6.07$)

Here in this work, we propose generalized MMs, for which MDs and more generally magnetic responses are not essential. It is revealed that higher-order electric resonances can totally substitute for MDs to obtain highly efficient MMs. Similarly according to the electromagnetic duality, we also show that higher order magnetic resonances can fully replace electric dipoles (EDs) to achieve electric mirrors (EMs).  We employ a simple while fundamental configuration of one-dimensional (1D) all-dielectric wire arrays to specifically demonstrate the generalized mirrors proposed, and further show that the electric resonance based MMs can outperform the conventional MMs relying on magnetic resonances. Our work can inspire more comprehensive investigations into higher-order multipolar excitations and full-angular interferences, and can spur lots of further investigations to achieve many other advanced functionalities replying on sole electric resonances, which were originally believed exclusive to magnetic responses.%in structures incorporating gain and other active, low-dimensional and topological materials.

The 1D infinite periodic configuration we employ is shown  in Fig.~\ref{fig1}(a): a plane wave is incident on the wires (for each wire: radius is $R$, inter-wire distance is $d$, and the refractive index is set to $n=3.5$ throughout this work) with wave-vector $\textbf{k}$ on the $\textbf{x}-\textbf{y}$ plane and its polar angle is $\phi$; it can be $\textbf{\textit{s}}$-($\mathbf{E}_0||\mathbf{z}$) or $\textbf{\textit{p}}$-polarized ($\mathbf{H}_0||\mathbf{z}$). We start with an individual wire, of which the normalized (by the single channel scattering limit $2\lambda/\pi$ and $\lambda$ is the wavelength in free space) scattering cross section is~\cite{Bohren1983_book,LIU_ArXivPrepr.ArXiv170406049_scattering_2017}: $N_{\rm sca}^{p,s} = \sum\nolimits_{m = -\infty}^\infty|a^{p,s}_m|^2$. Here $a_m^{p,s}$ are the scattering coefficients for $\textbf{\textit{p}}$- and $\textbf{\textit{s}}$- polarizations respectively, which can be calculated analytically~\cite{Bohren1983_book}; and the rotational symmetry of the wire leads to $a_m^{p,s}=a_{-m}^{p,s}$. We emphasize that~\cite{LIU_ArXivPrepr.ArXiv170406049_scattering_2017}: for $\textbf{\textit{p}}$-polarization, $a_{0:2}^{p}$ corresponds to the MD, ED, and electric quadrupole (EQ), respectively; while for $\textbf{\textit{s}}$-polarization, $a_{0:2}^{s}$ corresponds to the ED, MD, and magnetic quadrupole (MQ) respectively. In Figs.~\ref{fig1}(b) and (c) we show scattering spectra of a single wire for both polarizations, where the contributions from multipoles up to quadrupoles ($m=2$) are shown. Since $a_1^{s}=a_0^{p}$, it means that, despite the difference in magnitudes, for both polarizations MD resonances coincide spectrally as shown in Figs.~\ref{fig1}(b) and (c)~\cite{LIU_ArXivPrepr.ArXiv170406049_scattering_2017}.

Now it is clear that a simple wire can provide full sets of electromagnetic multipoles to play with. A detailed study shows that~\cite{Liu2014_ultradirectional}, in the backward and forward directions, multipoles of different natures (electric or magnetic) and/or orders show distinct scattering parities: the backward and forward scattering (in terms of electric field) is in-phase (even parity) for EDs and out-of-phase (odd parity) for MDs; more generally, the scattering parities are opposite: for multipoles of the same order but different natures due to electromagnetic duality (\textit{e.g.}, odd for EQ while even for MQ); and for multipoles of the same nature but neighbouring orders (\textit{e.g.}, even for ED while odd for EQ). The results are summarized schematically in Fig.~\ref{fig1}(d), where we assume the field is incident from the left with the electric field $\mathbf{E_0}$ (upward for ED and EQ, and out of the plane for MD and MQ). For passive multipoles, the forward scattering should be out of phase with respective to the incident field to generate destructive interference induced extinctions~\cite{Bohren1983_book}. As a result, \textit{in the backward direction}: the even parity of EDs leads to destructive interferences, which is the origin of EMs and also the half-wave loss; in contrast, the odd parity of MDs results in constructive interferences, based on which the MMs can be obtained. Naturally  from Fig.~\ref{fig1}(d) we expect that MQ and EQ can be employed to obtain  EMs and MMs respectively, as it is the scattering parity (odd for MMs and even for EMs), rather than resonance nature or order that can decide the types of mirrors.

Then we turn to the 1D periodic structure shown in Fig.~\ref{fig1}(a) to demonstrate what we propose.  The more general problem of normal scattering of plane waves by an arbitrary ensemble of parallel wires has been thoroughly studied using the multiple scattering method~\cite{FELBACQ_J.Opt.Soc.Am.AJOSAA_scattering_1994,LIU_ArXivPrepr.ArXiv170406049_scattering_2017}. For the $j^{th}$ wire (centred at $\mathbf{r}_j$ on the $x-y$ plane, and $j=1:N$) within such a scattering ensemble consisting $N$ wires, the scattered fields can be expressed with a sum of series of cylindrical harmonics, and the expansion (scattering) coefficients for the  $j^{th}$ wire ($a_{jm}^{s,p}$) are related to the single-wire scattering coefficients $a_m^{s,p}$ through ~\cite{FELBACQ_J.Opt.Soc.Am.AJOSAA_scattering_1994,LIU_ArXivPrepr.ArXiv170406049_scattering_2017}, $a_{jm}^{s,p}  + a_{jm}^{s,p} \sum\nolimits_{q \ne j}^{q = 1:N} {\sum\nolimits_{l =  - \infty }^\infty  {\Omega_{jm,ql} a_m^{s,p} } }  = e^{-im\phi+\mathbf{k}\cdot \mathbf{r}_{j} } a_m^{s,p}$, where $\Omega_{jm,ql}=i^{l - m} \mathbf{H}^{(1)}_{l- m} (k|\mathbf{r}_{qj}| )e^{i(m - l)\phi _{qj}}$ is the coupling matrix between the $m^{th}$ cylindrical harmonic of the $j^{th}$ wire and the $l^{th}$ cylindrical harmonic of the $q^{th}$ wire; $\mathbf{H}^{(1)}$ is the Hankel function of the first kind; $\mathbf{r}_{qj}=\mathbf{r}_{j}-\mathbf{r}_{q}$; and $\phi _{qj}$ is the polar angle of $\mathbf{r}_{qj}$.  For the infinite 1D periodic configuration we employ, the above equations can be further simplified by setting: $\mathbf{r}_j=jd\vec{\textbf{y}}$ ($j=-N:N,~N\rightarrow\infty$), $\phi _{qj}=\textbf{sgn}(j-q) \pi/2$ and according to the Floquet theory: $a_{jm}^{p,s}$=$a_{0m}^{p,s}e^{ik_yr_j}$, where $k_y=k\sin\phi$. Then we obtain~\cite{YASUMOTO_ElectromagneticTheoryandApplicationsforPhotonicCrystals_modeling_2005,BULGAKOV_Phys.Rev.A_bloch_2014}:
%--------------------------------------------------------------
\begin{equation}
\label{lattice_equation}
(\widehat{I}-\widehat{T}\cdot\widehat{C})\textbf{A}_0=\widehat{T}\textbf{B},
\end{equation}
%-------------------------------------------------------------
where $\widehat{I}$ is the identity matrix; $\widehat{T}_{ml}=-\delta_{ml}a_m^{p,s}$ ($\delta_{ml}$ is Kronecker delta function); $\textbf{A}_0=\{a_{0m}^{p,s}\}$; and $\textbf{B}=\{ e^{-im(\pi/2-\phi})\}$. The lattice sum matrix $\widehat{C}$ can be expressed as:
$\widehat{C}_{ml}=\sum\nolimits_{j = 1}^\infty  {H_{m - l}^{(1)} (jkd)[e^{ik_yj d}  + ( - 1)^{m - l} e^{ - ik_yj d} ]}$, the calculation of which is crucial and can be more efficiently solved through its integral form~\cite{YASUMOTO_ElectromagneticTheoryandApplicationsforPhotonicCrystals_modeling_2005,YASUMOTO_IEEETrans.AntennasPropag._efficient_1999}. Through solving Eq.~(\ref{lattice_equation}), the scattering coefficients $a_{0m}^{p,s}$ can be obtained. In a similar way, through implementing the boundary conditions, the field expansion coefficients inside the wires can be also obtained~\cite{YANG_IEEETrans.Geosci.RemoteSens._twodimensional_2005-1}£¬ and then all-space fields can be directly calculated.
\begin{figure}[tp]
\centerline{\includegraphics[width=8.9cm]{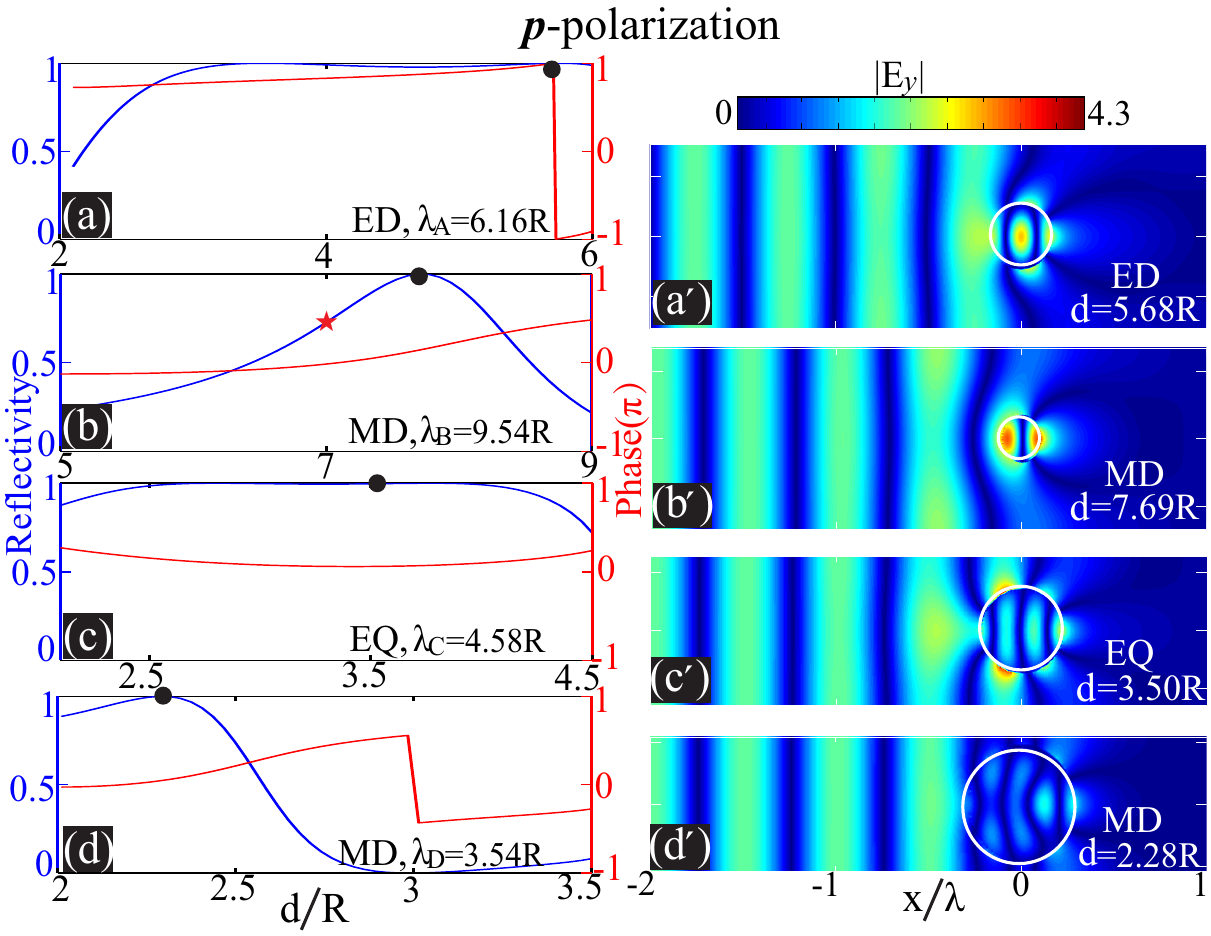}} \caption{\small (a)-(d) Reflectiviy and phase spectra with respect to $d$ at the four points A-D indicated in Fig.~\ref{fig1}(b) for $\textbf{\textit{p}}$-polarized incident waves. Four points with unit reflectivity and EM-type [$\alpha\approx\pi$, (a)] or MM-type [$\alpha\approx0$, (b)-(d)] phases are indicated by black dots and the corresponding field distributions by $|\textbf{E}_y|$ are shown in ($\rm a^\prime$)-($\rm d^\prime$), respectively. The red star indicates the point where there is perfect MM-type phase but relatively low reflectivity. White circles denote the wire boundaries, as is the case in Fig.~\ref{fig3}.}
\label{fig2}
\end{figure}

With the scattering coefficients obtained, the reflection coefficients in terms of electric fields ($r^p=\textbf{E}_y/\textbf{E}_{0}$ and $r^s=\textbf{E}_z/\textbf{E}_{0}$) for the $v^{th}$ diffraction order are~\cite{YASUMOTO_ElectromagneticTheoryandApplicationsforPhotonicCrystals_modeling_2005,BULGAKOV_Phys.Rev.A_bloch_2014}:
%--------------------------------------------------------------
\begin{equation}
\label{reflection_coefficients}
r^{p,s}_v(\phi)=\pm\frac{2}{d\sqrt{k_xk_{xv}}}\sum\limits_{m = -\infty}^\infty(-\frac{ik_{yv}+k_{xv}}{k})^ma_{0m}^{p,s},
\end{equation}
%-------------------------------------------------------------
where $v$ is the diffraction order; $k_{yv}=k_y+2v\pi/d$; $k_{xv}=(k^2-k_{yv}^2)^{1/2}$; and $k_x=k_{x0}=k\cos\phi$. Here in this work, we focus on the regime where there is only zeroth-order diffraction: $k^2-k_{yv}^2<0$ and $r^{p,s}_v=0$ for $|v|\geq1$. Then Eq.~(\ref{reflection_coefficients}) can be simplified as: $r^{p,s}_0(\phi)=\pm\frac{2}{dk_x}\sum\nolimits_{m = -\infty}^\infty(-\frac{ik_{y}+k_{x}}{k})^ma_{0m}^{p,s}$.  For the simplest case of perpendicular incidence $\phi=0$ and thus $k_y=0$, we have:
%--------------------------------------------------------------
\begin{equation}
\label{reflection_coefficients_simplified}
r^{p,s}_0(\phi=0)=\pm\frac{2}{dk_x}\sum\limits_{m = -\infty}^\infty(-1)^ma_{0m}^{p,s}.
\end{equation}
%-------------------------------------------------------------
The implications of the $\pm$ sign in Eq.~(\ref{reflection_coefficients})-(\ref{reflection_coefficients_simplified}) are clear: for the same $m$, the scattering coefficients for different polarizations would correspond to multipoles of the same order but of different natures and thus there is a $\pi$ phase difference between the electric fields reflected. Moreover, the $(-1)^m$ term in Eq.~(\ref{reflection_coefficients_simplified}) suggests that multipoles of the same nature but neighbouring orders would result in the same phase difference. This is consistent with the parity analysis of various multipoles shown in  Fig.~\ref{fig1}(d).

We begin with $\textbf{\textit{p}}$-polarized perpendicularly incident plane waves ($\phi=0$) and select four frequency points [as indicated in Fig.~\ref{fig1}(b): $\lambda_{\rm A}=6.16R$, $\lambda_{\rm B}=9.54R$, $\lambda_{\rm C}=4.58R$ and $\lambda_{\rm D}=3.54R$], where there are efficient ED, MD, EQ, and higher-radial number MD~\cite{LIU_ArXivPrepr.ArXiv170407994_superscattering_2017-1} resonance excitations within each wire, respectively. The reflection properties (dependence on $d$) for the whole periodic structure at those four points are summarized in Figs.~\ref{fig2}(a)-(d), where both the reflectivity ($\Re=|r^{p}_0(\phi=0)|^2$) and the reflection phase ($\alpha=\textbf{Ang}[r^{p}_0(\phi=0)]\in[-\pi,\pi]$) are shown. We note that throughout this work, we focus on the cases with near unit reflectivity ($\Re\approx1$). Other cases of perfect EM-type phase ($\alpha=\pi$) or MM-type phase [$\alpha=0$, such as the point indicated by the red star in Fig.~\ref{fig2}(b)] while of low reflectivity are not what we are interested in. Those types of reflection with low reflectivity are rather easy to obtain considering the normal reflection between two dielectric media: reflection from the denser (thinner) to the thinner (denser) side would exactly produce perfect EM-type (MM-type) phases according to the Fresnel equations.
\begin{figure}[tp]
\centerline{\includegraphics[width=8.9cm]{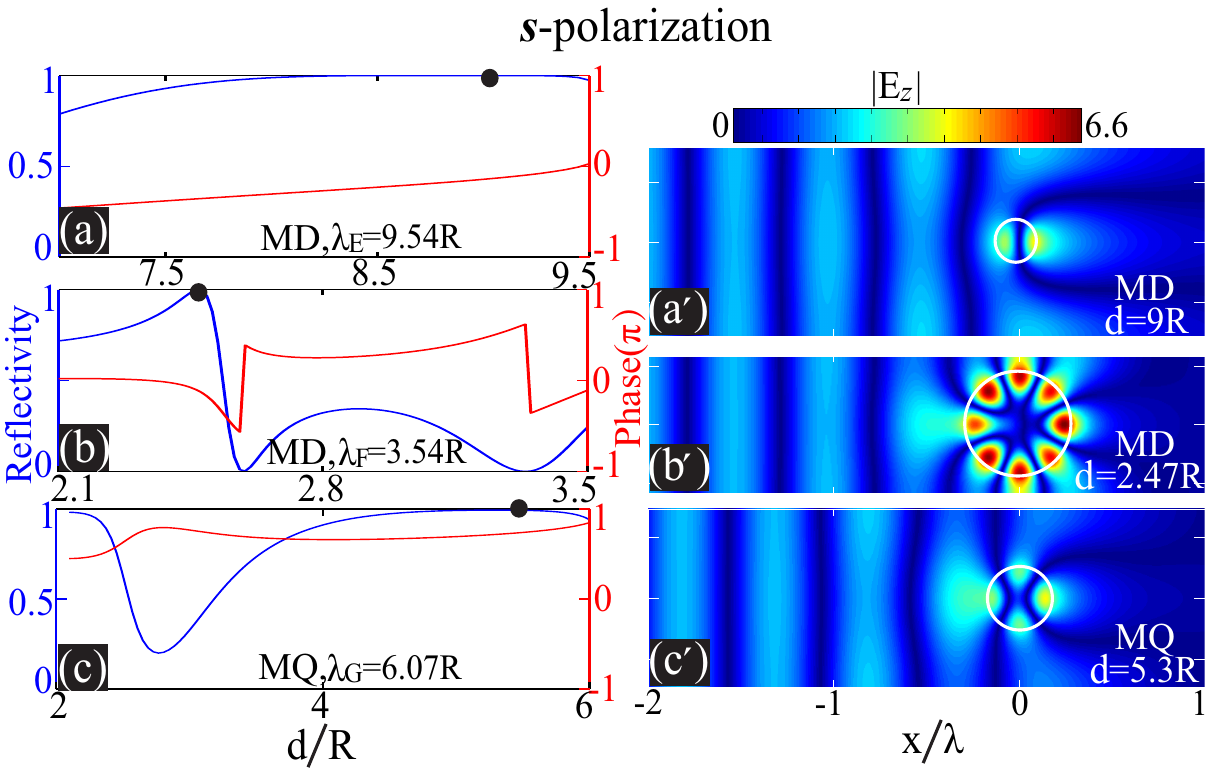}} \caption{\small (a)-(c) Reflectiviy and phase spectra with respect to $d$ at the three points E-G indicated in Fig.~\ref{fig1}(c) for $\textbf{\textit{s}}$-polarized incident waves. Three points with unit reflectivity and MM-type [(a) and (b)] or EM-type (c) phases are indicated by black dots  and the corresponding field distributions by $|\textbf{E}_z|$ are shown in ($\rm a^\prime$)-($\rm c^\prime$), respectively.}\label{fig3}
\end{figure}

The points of interest with unit reflectivity are indicated by black dots in Figs.~\ref{fig2}(a)-(d), which correspond to unit cell sizes of $d=5.68R,~7.69R,~3.5R,~2.28R$, respectively. It is clear that with efficient ED excitation, a near perfect EM is obtained [Fig.~\ref{fig2}(a), $\alpha\approx\pi$]; while when MD or EQ are efficiently excited, near perfect MMs have been obtained [Fig.~\ref{fig2}(b)-(d), $\alpha\approx0$]. In consistence with our former discussions with Fig.~\ref{fig1}(d), MDs or more generally magnetic multipoles are not indispensable for MMs, which as we show can be obtained relying on resonant electric multipolar excitations [Fig.~\ref{fig2}(c)]. Moreover, MMs can be obtained based on MDs of not only zero radial number~\cite{LIU_ArXivPrepr.ArXiv170407994_superscattering_2017-1} (which is widely employed for MMs~\cite{LIU_OpticaOPTICA_optical_2014,MOITRA_ACSPhotonics_largescale_2015,LIN_Appl.Phys.Lett._dielectric_2016,MA_ACSPhotonics_terahertz_2016-1,SONG_Conf.LasersElectro-Opt.2016Pap.JW2A27_silicon_2016}), as is shown in Fig.~\ref{fig2}(a), but also of higher radial numbers, as shown in Fig.~\ref{fig2}(d). To further demonstrate the different types of mirrors obtained at different multipolar excitations, we show correspondingly in Figs.~\ref{fig2}($\rm a^\prime$)-($\rm d^\prime$) the field distributions in a unit cell along $\textbf{y}$ (in terms of $|\textbf{E}_y|$, as far from the wire for $\textbf{\textit{p}}$-polarization, other electric field components are negligible; the axes are normalized by the wavelength).  It is clear that the nodes and anti-nodes positions of standing waves for MMs [Figs.~\ref{fig2}($\rm b^\prime$)-($\rm d^\prime$)] are shifted by $\bigtriangleup x\approx\lambda/4$ compared to those for the EM [Fig.~\ref{fig2}($\rm a^\prime$)], indicating exactly an $\bigtriangleup\alpha\approx\pi$ reflection phase difference. The most significant information is that, not as the name suggests, MMs do not necessarily need magnetic responses and can be realized based on sole electric ones [Figs.~\ref{fig2}(c) and ($\rm c^\prime$)].  We further note that our approach to obtain MMs in a thin layer of all-dielectric wires is fundamentally different from that of Ref.~\cite{ZHOU_Opt.Lett.OL_proposal_2016}. Rather than phototonic bandgaps, we rely on efficient multipolar excitations to produce sufficiently large extinctions to eliminate the transmissions~\cite{DU_Phys.Rev.Lett._nearly_2013,MIRZAEI_Phys.Rev.Lett._optical_2015}.
\begin{figure}[tp]
\centerline{\includegraphics[width=8.8cm]{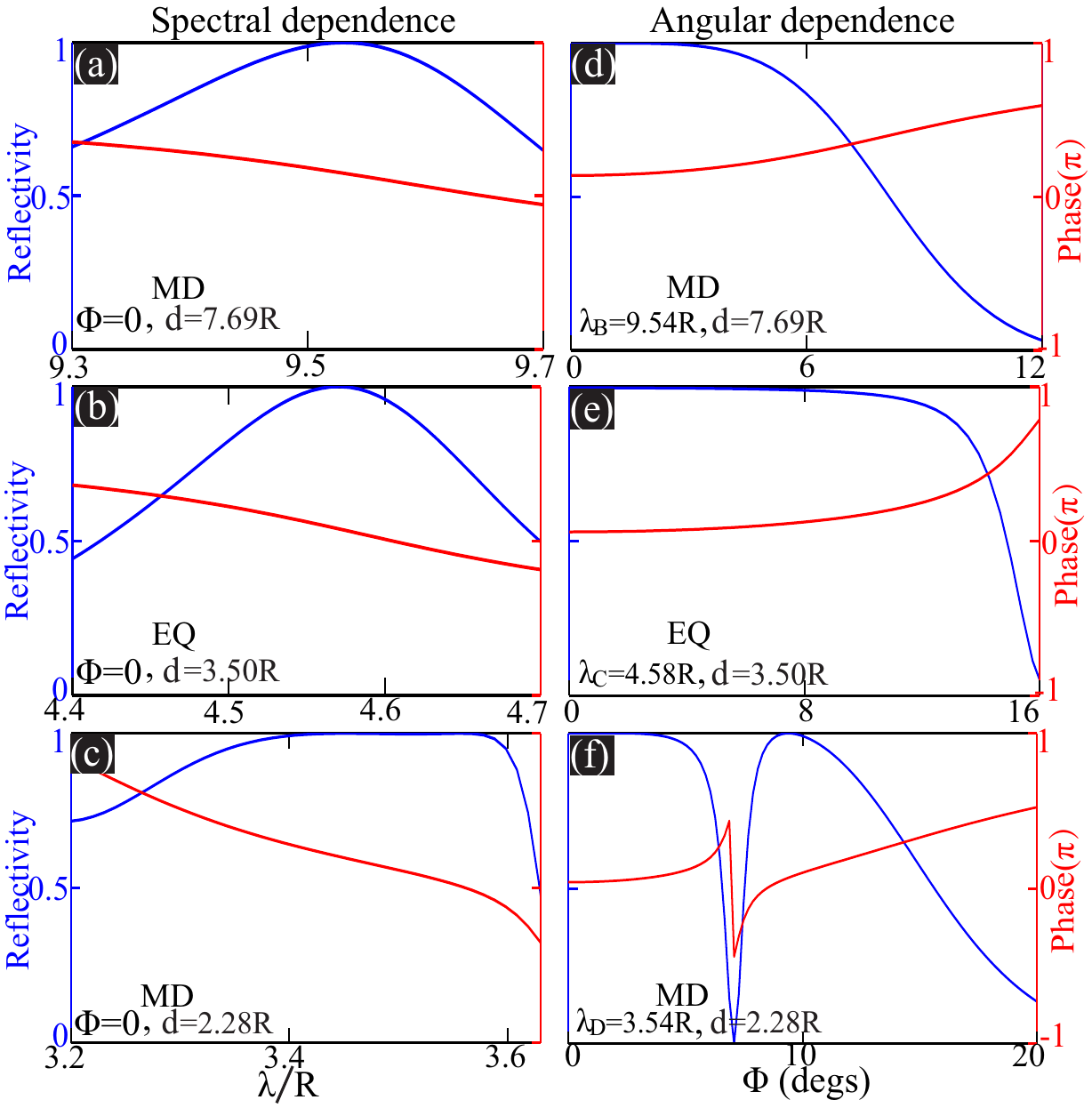}} \caption{\small Spectral (a)-(c) and angular (b)-(d) dependences of the reflectiviy and phase for the three MM points shown in Figs.~\ref{fig2}(b)-(d) with $\textbf{\textit{p}}$-polarized incident waves. For spectral dependence, we fix $d$ and $\phi=0$ and for angular dependence we fix $d$ and $\lambda$.}
\label{fig4}
\end{figure}

Then we switch to $\textbf{\textit{s}}$-polarized perpendicularly incident plane waves ($\phi=0$) and select three points, which are indicated in Fig.~\ref{fig1}(c): $\lambda_{\rm E}=9.54R$, $\lambda_{\rm F}=3.54R$, and $\lambda_{\rm G}=6.07R$, where there are efficient MD, MQ, and similarly higher radial number MD~\cite{LIU_ArXivPrepr.ArXiv170407994_superscattering_2017-1} resonance excitations, respectively. The reflectivity and phase spectra are shown in Figs.~\ref{fig3}(a)-(c), and the field distributions (in terms of $|\textbf{E}_z|$, which is the only non-zero electric field component for $\textbf{\textit{s}}$-polarization) at the indicated points of unit reflectivity ($d=9R,~2.47R,~5.3R$) are shown correspondingly in Figs.~\ref{fig3}($\rm a^\prime$)-($\rm c^\prime$). Here is clear that we have also obtained both MMs [Figs.~\ref{fig3}(a)-(b) and Figs.~\ref{fig3}($\rm a^\prime$)-($\rm b^\prime$), $\alpha\approx0$] and an EM [Figs.~\ref{fig3}(c) and ($\rm c^\prime$), $\alpha\approx\pi$]. Compared to the MM shown in Figs.~\ref{fig2}($\rm d^\prime$)-($\rm d^\prime$),  the difference at $\lambda_{\rm F}=3.54R$  [shown in Fig.~\ref{fig3}(b)] is that, despite the effective MD excitation, there are also significant ED [Fig.~\ref{fig1}(c)] and magnetic hexadecapolar excitations [Fig.~\ref{fig3}($\rm b^\prime$)]. Nevertheless, the reflection properties of the MM have been still preserved [Figs.~\ref{fig3} (b) and ($\rm b^\prime$)]. Similarly, not as its name suggests, EMs do not necessarily need electric responses, which can be also obtained by resonantly exciting  magnetic multipoles [Figs.~\ref{fig3}(c) and ($\rm c^\prime$)]. In a word, we cannot simply judge the types of mirrors obtained solely through the nature of the modes excited. But rather both the nature and the order of excitations (which together decide the scattering parity) should be considered.

Up to now, we have confined our investigations to fixed frequencies (though for dielectric structures the results obtained are fully scalable) and incident angle ($\phi=0$). As a next step, we investigate the spectral [Figs.~\ref{fig4}(a)-(c)] and angular dependences [Figs.~\ref{fig4}(d)-(f)] of the MMs shown in Figs.~\ref{fig2}(b)-(d). Basically the MMs are based on resonant excitations [Figs.~1(b)-(c)] and thus can function only in a relatively narrow spectral regimes [Figs.~\ref{fig4}(a)-(b)] for both effective MD (of zero radial number, Fig.~\ref{fig4}(a)] and EQ [Fig.~\ref{fig4}(b)] excitations. In contrast, the MM obtained at MD of higher radial number is effectively broadband [almost an achromatic MM obtained, Fig.~\ref{fig4}(c)], since for a single wire there is correspondingly a flat-banded dispersion curve for the MD [see the region close to point D in Fig.~\ref{fig1}(b)].  For angular dependence, the MM obtained based on EQ [Fig.~\ref{fig4}(e)] is superior to those based on MDs [Fig.~\ref{fig4}(d) and (f)]. This means that the such a generalized MM based on an electric resonance can function over  wider incident angle ranges or for relatively narrower incident beams with higher spatial harmonics.

In conclusion,  we propose the concept of generalized MMs, which we demonstrate can be realized through exciting only electric resonances. Within a simple single-layered all-dielectric wire arrays, we show that higher-order electric resonances can fully replace magnetic ones for MM generations, and similarly higher order magnetic resonances can substitute electric ones for EM generations. Moreover, for angular performances we show that generalized MMs based on electric resonances can be superior to the conventional MMs based on magnetic resonances. Here we confine our discussions to multipoles up to quadrupoles and generalized EMs and MMs can of course be achieved replying on much higher-order resonances. The principles we reveal are quite universal and can be applied within other configurations consisting of spherical~\cite{Supplemental_Material} or other irregularly shaped particles made of dielectric, plasmonic (including plasmonic holes~\cite{ESFANDYARPOUR_Nat.Nanotechnol._metamaterial_2014,ROTENBERG_Phys.Rev.Lett._plasmon_2012,ROTENBERG_Phys.Rev.B_magnetic_2013}), two-dimensional, topological materials or their combinations. We believe that our work can stimulate lots of further studies aiming to obtain, replying on sole electric responses, many other advanced functionalities that were originally believed unobtainable without magnetic responses. This can hopefully simply a lot of structures and devices that are complex due to the requirement of  magnetic resonance excitations. At the same time, the proposed concepts can probably trigger more comprehensive investigations into higher-order multipolar excitations (involving electric, magnetic and toroidal multipoles~\cite{LIU_ArXivPrepr.ArXiv160901099_multipolar_2016,PAPASIMAKIS_NatMater_electromagnetic_2016}), full-angular interferences and more flexible phase control with passive and/or active materials~\cite{PARK_NanoLett._dynamic_2017,JAHROMI_ACSPhotonics_transparent_2017}, which may provide new initiatives for advanced fundamental studies and incubate lots of novel applications based on manipulations of light-matter interactions.

%\section*{Acknowledgments}
We thank Y. Xu, A. E. Miroshnichenko and Y. S. Kivshar for initial inspiring discussions, and acknowledge the financial support from the National Natural Science Foundation of China (Grant number: $11404403$), and the Outstanding Young Researcher Scheme of National University of Defense Technology. The author would like to dedicate this work to his wife Sijia Yang, and his newborn daughter Qingyang (Eileen) Liu.

%\section*{References}
%\bibliographystyle{osajnl}
%\bibliography{References_scattering}

%==========================
\end{document}